\documentclass[twoside,12pt]{article}
\usepackage{style}

\usepackage{enumerate}
\usepackage{multibib}
\newcites{supp}{Supplementary References}

\usepackage[utf8]{inputenc}
\usepackage{mathtools}

\usepackage{setspace}
\usepackage{dsfont}
\usepackage{epstopdf}
\usepackage{boxedminipage}
\usepackage{amsfonts}
\usepackage{setspace}
\usepackage{natbib}
\usepackage{bm}
\usepackage{wasysym}
\usepackage{textcomp}
\usepackage[export]{adjustbox}
\usepackage{mathtools} 
\usepackage[normalem]{ulem}
\usepackage{setspace,lscape}
\usepackage{tabularx,booktabs,multirow}
\usepackage[usenames]{color}
\usepackage{soul}

\usepackage{color,latexsym,epsfig,graphics,graphicx,amssymb,amsbsy,amsmath}
\usepackage{rotating}
\usepackage[arrow,curve,all]{xy}
\usepackage{setspace,lscape}
\usepackage{tabularx,booktabs,multirow}
\usepackage{tikz}
\usetikzlibrary{shapes.callouts}
\usetikzlibrary{snakes}

\newcommand{\cblue}{\color{black}}

\newcommand{\beq}{\begin{equation}}
\newcommand{\eeq}{\end{equation}}
\newcommand{\beas}{\begin{eqnarray*}}
\newcommand{\eeas}{\end{eqnarray*}}
\newcommand{\bea}{\begin{eqnarray}}
\newcommand{\eea}{\end{eqnarray}}
\newcommand{\bei}{\begin{itemize}}
\newcommand{\eei}{\end{itemize}}
\newcommand{\ben}{\begin{enumerate}}
\newcommand{\een}{\end{enumerate}}
\newcommand{\bet}{\begin{theorem}}
\newcommand{\eet}{\end{theorem}}
\newcommand{\bel}{\begin{lemma}}
\newcommand{\eel}{\end{lemma}}
\newcommand{\bep}{\begin{proposition}}
\newcommand{\eep}{\end{proposition}}
\newcommand{\bed}{\begin{definition}}
\newcommand{\eed}{\end{definition}}
\newcommand{\bec}{\begin{corollary}}
\newcommand{\eec}{\end{corollary}}
\newcommand{\bex}{\begin{example}}
\newcommand{\eex}{\end{example}}

\def\0{\boldsymbol{0}}

\newenvironment{boxedtable}[2]{\begin{table}[!htbp]\noindent{\caption{#1 \label{#2}}}\small} {\noindent\end{table}}

\newcommand{\tablexplain}[1]{\noindent\small{#1}}
\newenvironment{ctabular}{\begin{center}\begin{tabular}}{\end{tabular}\end{center}}

\newcolumntype{Y}{>{\centering\arraybackslash}X}

\usepackage{lastpage}
\usepackage{url}

\ShortHeadings{Ranking by Lifts (RBL) by Basu and Berman}{Ranking by Lifts (RBL) by Basu and Berman}
\firstpageno{1}

\begin{document}




\title{Ranking by Lifts:~A Cost-Benefit Approach to Large-Scale A/B Tests}



       
       \author{\name Pallavi Basu  \email pallavi\_basu@isb.edu \\
       \addr Operations Management\\
       Indian School of Business\\
       Hyderabad, Telangana 500111, India
       \AND
       \name Ron Berman  \email ronber@wharton.upenn.edu \\
       \addr The Wharton School\\
       University of Pennsylvania\\
       Philadelphia, PA 19104, USA}

\date{\today}

\maketitle \thanks


\begin{abstract}

A/B testing is a core tool for decision-making in business experimentation, particularly in digital platforms and marketplaces. Practitioners often prioritize lift in performance metrics while seeking to control the costs of false discoveries. This paper develops a decision-theoretic framework for maximizing expected profit subject to a constraint on the cost-weighted false discovery rate (FDR). We propose an empirical Bayes approach that uses a greedy knapsack algorithm to rank experiments based on the ratio of expected lift to cost, incorporating the local false discovery rate (lfdr) as a key statistic. The resulting oracle rule is valid and rank-optimal. In large-scale settings, we establish the asymptotic validity of a data-driven implementation and demonstrate superior finite-sample performance over existing FDR-controlling methods. An application to A/B tests run on the Optimizely platform highlights the business value of the approach.
\end{abstract}










\section{Introduction}


A/B testing is often used to optimize websites, marketing campaigns, and user interfaces. Leading companies report running over 10,000 experiments annually \citep{kohavi2017surprising}. 
Recently, academics introduced new extensions to the 
statistical tool of basic hypothesis tests that fits the practicalities of A/B tests in the field. Examples include computing profit-maximizing sample sizes \citep{feit2019test}, designing experiments when samples are limited \citep{azevedo2020b}, and controlling for peeking behavior and optional stopping \citep{johari2022always}. One concern when running many tests is multiple hypothesis testing, which results in false discoveries. Common solutions to control for false discoveries include the Benjamini-Hochberg (BH) \citep{benjamini1995controlling}, Sun and Cai (SC) \citep{sun2007oracle}, and the weighted false discovery rate control (BCDS) \citep{basu2018weighted} methods. 

{\cblue We base our framework on a prevalent setup in the literature \citep{azevedo2020b, sudijono2024optimizing} where a collection of ideas, often having a common connection, are tested using limited resources. The authors in the latter work call the setup `experimentation programs'. In these programs, many A/B tests are run concurrently, and each test often compares} a new intervention to a standard practice baseline used in production. For example, comparing a new recommendation algorithm to an existing one \citep{kim2019champion, goli2024bias}, or trying a new targeting policy compared to an existing one \citep{simester2020efficiently}. These experiments are sometimes called champion-challenger experiments, and their uniqueness is that the experimenter has long-term observations about the average outcomes of the baseline treatment. Further, if many such experiments are being run, the cost of switching from the champion to the challenger might differ for each experiment.

In this paper, we develop a false discovery control method that focuses on lifts (i.e., a transformation of Relative Risks) while considering the available long-term information about baseline outcomes and the potential differences in the costs of implementing the winning intervention. We take a cost-benefit approach to experimentation, aiming to maximize lift, a metric practitioners are often interested in, while controlling for acceptable costs that accrue from false discoveries.

Methodologically, we develop an oracle approach that yields a greedy knapsack algorithm and derive statistically feasible estimators to estimate the oracle statistic. Analyzing lifts in A/B tests is similar to relative risk analysis, prevalent in medical studies. We contribute to developing the greedy optimal oracle solution for the objective function involving lifts while considering false discovery control. In the context of relative risks, we also derive a second-order logarithm correction for the mean. 

We use a simulation study to demonstrate the gain in performance over existing state-of-the-art FDR control methodologies. Finally, we use data from 2,766 A/B tests conducted using a large online A/B testing platform (Optimizely) to compare the performance of our method to the other state-of-the-art methods.
We hope practitioners of A/B testing will find our method useful when their objective is to optimize lifts in Champion-Challenger experiments, subject to cost control.

\subsection{False Discovery Rate Control in A/B Testing}
In large-scale experiments that test many hypotheses simultaneously, false discovery rate (FDR) control is often recommended to provide more statistical power over the family-wise error rate (FWER), which controls the probability of having at least one false positive but does not control the expected false discovery proportions. \cite{berman2022false} reported up to 25\% two-sided error proportions when practitioners do not correct for multiple testing at 5\% significance. The inflation in error proportion is alarming and can lead to considerable wastage by switching to null interventions with implementation costs.  
Although false discoveries by themselves might not have a cost, if each test is weighted by the implementation costs, then the expected value of the ratio of false discoveries to the number of discoveries made (which is the definition of FDR) can be viewed as a proportion of `wasted' investment resources.
This phenomenon is more severe when the proportion of true null effects is higher. In our simulations, we find that for 80\% null effects, up to 50\% discoveries are false (one-sided) when not appropriately corrected at 5\% significance.

Our approach to FDR control is based on a line of work by \cite{sun2007oracle}, \cite{efron2008microarrays}, and \cite{basu2018weighted}, which relies on empirical Bayes estimation and adjustments, leading to a more robust methodology that also considers variations of power as an objective function to maximize discoveries.


\subsection{Relative Risk and Lift for Two-Sample t-Tests}

Relative risk has been a common metric of interest in epidemiology and medical research. It is often not recommended as the sole metric of interest but as part of an evaluation in conjunction with absolute risk or risk difference. In the context of A/B testing, relative risk surfaces through the interest of companies in comparing their experiments using lifts (which are the relative risks minus one). In this work, we study the impact of ranking multiple A/B test results by their expected lifts to decide which ones to implement. However, we anticipate our work to be helpful in epidemiological studies, economics, finance, and marketing. Irrespective of the domain of expertise, we refer to the users of A/B tests as experimenters and assume that the data has been generated from randomized trials.

\subsection{Executive Summary} The practical implications from our work provide the following recommendations to A/B testing practitioners:~
\begin{itemize}

\item An FDR control approach, such as BH or SC \citep{benjamini1995controlling, sun2007oracle} is recommended to avoid inflated false discoveries when testing multiple A/B tests. The correction is necessary even if the experiment has a second stage, replicating only the best variation in each test and following a first stage with several variations in each experiment. 

\item If experimenters are interested in using lifts as the primary metric, it is recommended that they do so by applying a logarithm transformation to the relative risk to ensure asymptotic normality. Further, a second-order mean correction of half the estimated variance is necessary to de-bias the conventional estimated mean. The p-value or t-statistic from this process can then be used for FDR control.

\item To accommodate the heterogeneity in baseline profits, we derived an optimal oracle testing procedure via a greedy knapsack approach, which we call Ranking by Lifts (RBL). In our data-driven application, and in practice, the empirical performance in FDR control validity matches that of the BCDS \citep{basu2018weighted} weighted multiple testing procedure while often providing a considerable power improvement. We recommend using our proposed RBL procedure when profit considerations are important since they can potentially provide a sizeable gain in net profit over conventional FDR control methods such as standard BH, SC, or BCDS.
\end{itemize}

\subsection{Paper Organization}
We define and formulate the experimenter's decision problem in Section \ref{sec:problem}. Section \ref{sec:oracle} describes the general optimization framework, optimal ranking, the conditional expectation of the lifts, and the main algorithm. We dedicate Section \ref{sec:est} to theoretical justifications about the validity and optimality of the oracle procedure, the second-order mean correction, and finally, the asymptotic validity of the data-driven procedure. We demonstrate the finite sample performance of the data-driven procedure in Section \ref{sec:sims} using simulation, while Section \ref{sec:app} illustrates an application to actual experiments conducted using the Optimizely platform. Section \ref{sec:related} discusses related work, and \ref{sec:disc} concludes with a discussion. The R codes for implementing the proposed method are available upon request.

\section{Problem Formulation}
\label{sec:problem}
An experimenter conducts $N$ experiments indexed by $i=1 \ldots N$. Each experiment contains a baseline existing treatment $i0$ called the ``champion'' and a new treatment called a ``challenger'' $i1$. We assume that switching from the champion to the challenger costs $c_i>0$. The decision maker's goal is to decide which experiments to switch from the champion to the challenger while ensuring that the proportional cost of erroneous switches (i.e., switching to the challenger when the champion is better) is less than $\alpha$.

The baseline conversion rate is $p_{i0}$ while the challenger's is $p_{i1}$. We denote the lift as $\ell_i = p_{i1}/p_{i0}-1$. The decision maker's switching decision will be coded as $\delta_i \in \{0,1\}$ when zero indicates staying with the champion, and one indicates switching. We denote $\theta_i = \mathbf{1} \{ \ell_i > 0 \} $ as the indicator for a positive true lift in experiment $i$, i.e., that switching to the challenger will be a correct decision. A quantity we will use to base the decision on is called the local false discovery rate (lfdr) and is defined as $Pr(\theta^{ts}_i=0|data)$ for two-sided decisions where $\theta^{ts}_i = \mathbf{1} \{ \ell_i \neq 0 \} $.

The objective of the experimenter is  to select values for $\delta_i$ to maximize $\mathbb{E} \sum_{i} \Pi_i \ell_i \delta_i$, subject to the constraint $\mathbb{E} \{ \sum (1-\theta_i) \delta_i c_i / \sum \delta_i c_i \} \leq \alpha$. The value $\Pi_i$ indicates the profit the firm makes from sending the champion treatment to the customers, including the conversion rate and the profit margin, which is multiplied by the number of customers. Hence, switching to the challenger will add an incremental profit from the improvement in conversion rates.
The expression for the constraint indicates that the expected proportional cost of erroneous switching decisions is bounded by $\alpha$. {\cblue This formulation allows the experimenter to trade off longer-term learning about interventions that have true effects, with maximizing short-term net-profit that can reject too many or too few hypotheses.\footnote{Appendix \ref{app:alternate} analyzes an alternate net-profit objective and provides more clarity behind the rationale for our objective of choice.}}
Each experiment has sample sizes $n_{i0}$ and $n_{i1}$ and observed conversions $Y_{i0}$ and $Y_{i1}$. The experimenter computes the estimated conversion rates $\widehat{p}_{i0} = Y_{i0}/n_{i0}$ and similarly for $i1$, yielding the estimated lifts $\widehat{\ell}_i = \widehat{p}_{i1}/\widehat{p}_{i0}-1$.

The decision maker must decide which treatments to implement based on the estimated lifts. We assume that the statistic $Z_i=\ln \widehat{p}_{i1}/\widehat{p}_{i0}$ is drawn from a mixture distribution 
$\pi_0 F_{i0} + (1 - \pi_0) F_{i1}$ where $\pi_0$ denotes the probability of the statistic being from a null lift, and $F_{0i}$ and $F_{1i}$ denote the CDFs of the observed statistics $Z_i$ under the null and the non-null respectively. As an example, when the null is true, then $\ln(p_{i1}/p_{i0})=0$, and if the observed statistics follow a normal distribution, then $F_{i0}$ is the cdf of $\mathcal{N}(0,\sigma_i^2)$. At the same time, when the alternative is true, then the conditional $F_{i1}$ is the cdf of $\mathcal{N}(\ln(p_{i1}/p_{i0}),\sigma_i^2)$ for some $\ln(p_{i1}/p_{i0}) \neq 0$.

To summarize, the experimenter solves:
$$
\max_{\delta_i \in \{0,1\}} \mathbb{E} \sum_{i} \Pi_i \ell_i \delta_i \text{~~subject to~~} \mathbb{E} \{ \sum (1-\theta_i) \delta_i c_i / \sum \delta_i c_i \} \leq \alpha.$$

\section{Oracle Procedure for Optimizing Lifts}
\label{sec:oracle}
We develop a greedy knapsack algorithm that takes the local false discovery rates (lfdr) as known and finds the optimal set of hypotheses to include in the knapsack (including refers to rejecting a null hypothesis) while maximizing the total profit in the knapsack subject to the cost constraints. This algorithm will require every hypothesis to be characterized by two components: its value and its weight. These components will correspond to the conditional expectations of  $\mathbb{E} [\Pi_i \ell_i|data]$ (value) and $\mathbb{E} [c_i (1-\theta_i -\alpha)|data]$ (weight) whose summation comprise the objective and constraint function components in Section \ref{sec:problem}. We first derive the expressions for these components and then describe the algorithm. Section \ref{sec:est} then provides statistical properties and details of the estimation procedure.

\subsection{Deriving the Value and Weight Components}

We first derive an expression that connects the lfdr to $\mathbb{E} [\Pi_i \ell_i|data]$.
Let $\zeta_i := \ln p_{i1} / p_{i0}$. Therefore from the definition of $\ell_i$,  we can rewrite  $\ell_i = e^{\zeta_i} - 1$, making the objective function $\sum_i \Pi_i \delta_i E [e^{\zeta_i} - 1 | data]$ where we used the fact that $\delta_i$ is a function of the data (and potential randomization which we will discuss later) to take it out of the conditional expectation. 

Following Equation 2.7 in \cite{romano2005testing}  which we modify using the conventions of \cite{efron2011tweedie}, we define a natural exponential family of distributions $f_{\zeta}(z) = C(\zeta) e^{\zeta z} f_0 (z)$. Invoking Bayes' rule, the posterior density of $\zeta|z$ is $g(\zeta|z) = f_{\zeta}(z) g(\zeta)/f(z)$, where $f(z)$ is the marginal density and $\zeta \sim g(\cdot)$ is a one-group model with possible atoms. We then rewrite $g(\zeta|z) = C(\zeta) g(\zeta) e^{\zeta z - \ln(f(z)/f_{0}(z)) }$. 

We rename $A(z) := \ln(f(z)/f_{0}(z))$, noticing that the $lfdr$ is $\pi_0 e^{-A(z)}$. The moment generating function (MGF) of $\zeta | z$ can be written as $e^{A(z+1) - A(z)}$, which we rewrite as 
$lfdr(z)/lfdr(z+1)$ for convenience of estimation later.
Recalling that the quantity of interest to us in the objective function is $E [e^{\zeta} - 1 | z]$, then $E [e^{\zeta} - 1 | z] = lfdr(z)/lfdr(z+1) - 1$. The weight component $\mathbb{E} [c_i (1-\theta_i -\alpha)|data]$ is connected to the lfdr by the expression $c_i(lfdr_i-\alpha)$.

\begin{figure}
\begin{center}
\includegraphics[scale=0.6]{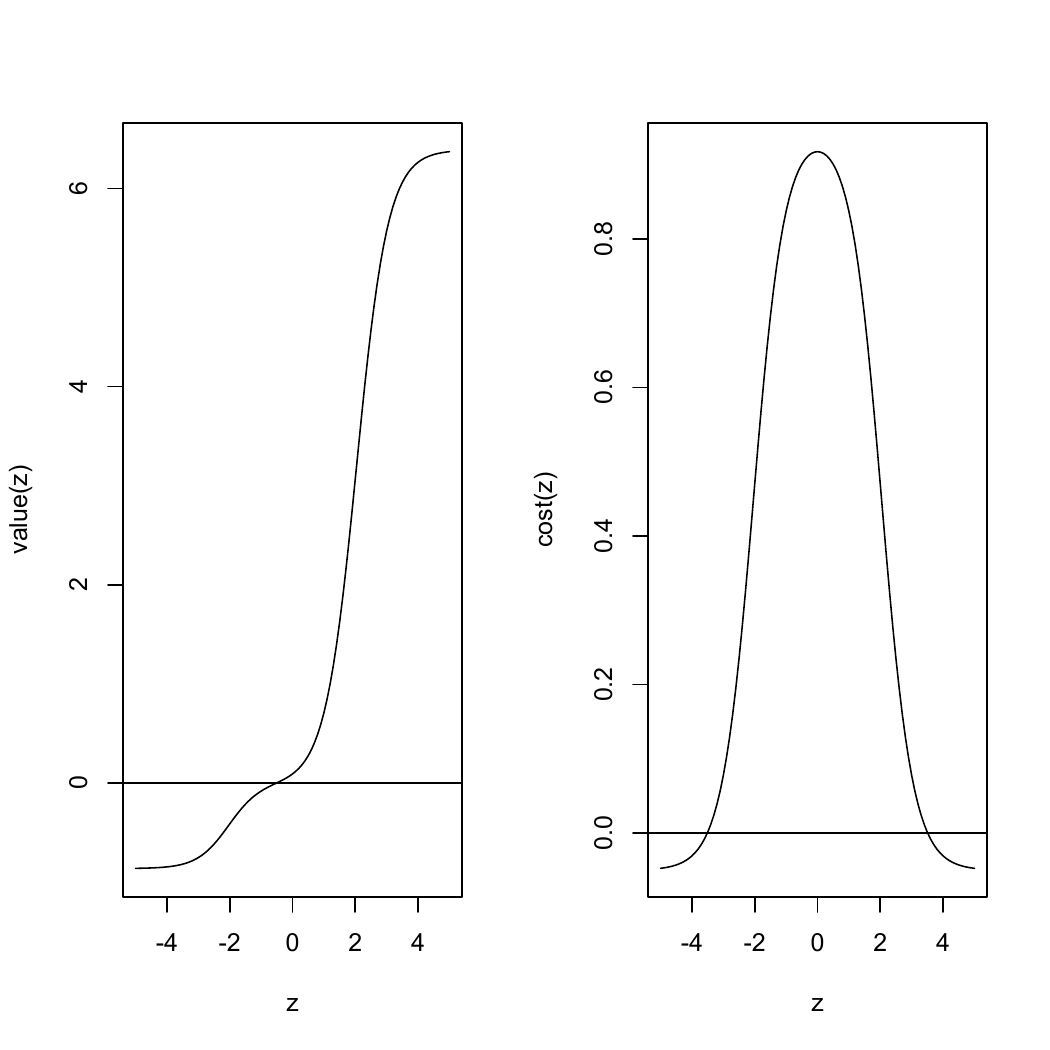}
\end{center}
\caption{Illustration of the value and the cost functions.}
\label{fig:plot_value_wt}
The figure displays the value and cost functions for z-values ranging from [-5, 5]. The underlying data-generating process is assumed to be from a normal mixture with a 0.8 probability of being the null of N(0, 1) and non-null distributions of N(2, 1) and N(-2, 1) with equal remaining mass. We choose $\alpha = 0.05$ for computing the cost. We note that roughly at $z = -4$, both the value and cost are negative; at $z = -2$, the value is still negative, and the cost becomes positive; at $z = 2$, the value turns positive while the cost stays positive, and at $z = 4$ the value stays positive, and the cost decreases below zero.
\end{figure}

Figure \ref{fig:plot_value_wt} provides a graphical illustration of the value and weight components for a mixture distribution where $F_0$ is $N(0,1)$ and $F_1$ is a mixture of $N(-2,1)$ and $N(2,1)$ with equal probability. The cost function is symmetric around 0 for different values of $z$. The value function is not symmetric. For very negative values of $z$, the value is negative. However, these hypotheses would contribute a positive value to the objective function for a mildly negative value of $z$ (roughly half of the standard deviation on the left of the vertical axis at zero). 

Remark: The definition of the lfdr we used assumes that $Z \sim \mathcal{N}(0,1)$ is symmetric under the null. In our case, some experiments have negative lifts that will generate low lfdr, although they should not be rejected (as they have a high weight). To accommodate that, we adjust the null density for $Z$ to be evaluated at $\sup_{\zeta \leq 0} N(\zeta, 1)$ which is essentially the same as $N(max(z, 0); 0, 1)$. Note that we do not use this adjustment for the `value' statistics but only for the `weight' statistics.\footnote{The null probability $\pi_0=Pr(\zeta \leq 0)$ might also be estimated as $ Pr(\zeta = 0) +  (1 - Pr(\zeta = 0))/2$ but, in our numeric experiments, we did not find sufficient practical difference in power and validity.}

\subsection{The Large Sample Distribution of $Z = \ln \hat{p}_{1}/\hat{p}_{0}$}
In the previous derivation, we assumed a large enough sample for $Z_i$ that facilitates the parametric exponential family distribution assumption of $f_{\zeta}(\cdot)$. This approach is popular in medical studies \citep{katz1978obtaining}, and in this Section, we provide details to justify this assumption.
The asymptotic distribution of $Z_i$ when  $n_{0i}, n_{1i} \to \infty$ is:
\[
Z_i = \ln \hat{p}_{i1}/\hat{p}_{i0} \approx N( \ln(p_{i1}/p_{i0}), \sigma_i^2),
\]
with $\widehat{\sigma_i^2} = S_i^2 := (1-\widehat{p}_{i1})/Y_{i1} + (1-\widehat{p}_{i0})/Y_{i0}$. Under the null, $p_{i1} = p_{i0}$; hence, the Normal distribution is centered at zero.
We recommend at least one hundred observations from each population.  

To derive the lfdr statistic using $Z_i$, we define 
$H_i:= Z_i/\sigma_i$ and $\eta_i = \zeta_i/\sigma_i$.\footnote{Normalizing $Z_i$ by $\sigma_i$ makes the statistics homoscedastic but does not alter the null.}
Recalling that our objective function is $E[e^{\zeta_i} - 1 | z_i]$, we can rewrite it as $E [e^{\sigma_i \eta_i} - 1 | h_i] = lfdr(h_i)/lfdr(h_i + \sigma_i) - 1$.


\subsection{The Proposed Greedy Knapsack Algorithm}

We now describe the main algorithm. Table \ref{tb:decision} provides a classification of the hypotheses based on their values and weights and the decision rule that will be used to determine if they are included in the knapsack (i.e., the null is rejected) or excluded (i.e., the null is not rejected).

Suppose the object's weight is non-positive and the value is positive. Hence, it is optimal always to include the hypotheses in quadrant I, and we include the hypothesis (rejecting the null). In that case, it does not reduce the capacity available in the knapsack and increases its value. 
Similarly, for those in quadrant III, the hypotheses do not add value and take up space (add cost) in the knapsack and hence should be excluded.

For the hypotheses with either negative weight and negative value (quadrant IV) or positive weight and positive value (quadrant II), we need to decide whether to include or exclude them based on the ratio of their value to weight. We would like to include those that either provide a high positive value to positive weight ratio (thus not taking too much space but adding a lot of value) or those with a low negative value to negative weight (therefore not reducing the total value in the knapsack by much, but adding capacity to the knapsack). The order of inclusion or exclusion is done by the rank of the value-to-weight ratio, which we prove is optimal in Section \ref{sec:opt}. The details of the algorithm are described below.


\begin{table}[ht]
\vspace{0.1in}
\begin{center}
\caption{Decision rule used to include or exclude hypotheses from the Knapsack}
\label{tb:decision}
\begin{tabular}
{|c|c|c|}\hline  Decision Rule & Weight $>$ 0 & Weight $<$ 0\\ \hline Value $>$ 0 & II:~Exclude and rank to decide & I:~Always reject the null\\ \hline Value $<$ 0 & III:~Always accept the null & IV:~Include and rank to decide \\ \hline
\end{tabular}
\end{center}
The algorithm will a) always include in the knapsack (reject the null) hypotheses in quadrant I, b) always exclude (fail to reject) in quadrant III, and c) decide on hypotheses in quadrants II and IV according to their ranking of value-to-weight ratio.
\end{table}


\vspace{0.1in}
\begin{center}
\fbox{\begin{minipage}{38em}
\textbf{Ranking by Lifts (RBL) Algorithm}

\begin{description}
    
\label{alg:greedy}
 \item Step 1. For each hypothesis compute the weight $c_i(lfdr_i(h_i)-\alpha)$, and the value $\Pi_i(lfdr_i(h_i)/lfdr_i(h_i + \hat{\sigma_i}) - 1)$.
  Initially, the knapsack is empty with a capacity of zero. First, all hypotheses with negative weight are added to the knapsack. This will enlarge the knapsack and give it a maximum capacity of \[\sum_{i:lfdr_i<\alpha} c_i(\alpha-lfdr_i).\]
 
 \item Step 2. Discard all hypotheses with positive weight and negative value from consideration, as they will take up knapsack space and reduce the total value of items in the knapsack.

 \item Step 3. Create a consideration set of the items with either negative weight and value or positive weight and value, and rank them based on the value-to-weight ratio \[\frac{\Pi_i(lfdr_i(h_i)/lfdr_i(h_i + \hat{\sigma}_i) - 1 )}{c_i(lfdr_i(h_i) - \alpha)}.\]

 \item Step 4. For the items in the consideration set, start with the highest-ranked item, and for each item:
 
 (i) If it has a negative weight and value (which means it was added in Step 1), discard it from the knapsack.
 
 (ii) Add it if it has positive weight and value (which means it is currently outside the knapsack).

 Stop when there is no more knapsack capacity available.




\end{description}

\end{minipage}}
\end{center}
\vspace{0.2in}

\section{Statistical Guarantees}
\label{sec:est}
First, we show the properties of the oracle procedure that assumes knowledge of $\pi_0$, $F_0$, and $F_1$. We then discuss the second-order mean correction and the estimation of the local FDR statistics, which are used in the data-driven implementation of the algorithm. We conclude by analyzing the data-driven procedure's asymptotic false discovery rate control guarantee.

\subsection{Properties of the Oracle Procedure}
\label{sec:opt}

\vspace{2mm}
\noindent
{\bf Proposition 1 (Validity.)} {\it The proposed oracle procedure controls the FDR at the desired level.} 


\vspace{2mm}
\noindent
{\bf Proof}.
The proof follows by noting that the design of the procedure is such that we control the conditional expectation, i.e., $E(\sum (1-\theta_i)\delta_ic_i/\sum\delta_ic_i|data) \leq \alpha$. This is done by stopping the algorithm from adding additional hypotheses to the knapsack when no more capacity is available. Since we control the conditional expectation, the unconditional expectation, too, is controlled at the desired threshold. 
\hfill\BlackBox

\vspace{5mm}
\noindent
{\bf Proposition 2 (Optimality.)} {\it The proposed oracle procedure optimally ranks the hypotheses, with some decisions being exactly optimal regardless of the ranking.}

\vspace{2mm}
\noindent
{\bf Proof}.
We first explain what the statement entails and then prove it. The statement says that decisions in quadrants I and III of Table \ref{tb:decision} are exact-optimal. We demonstrate that starting from any decision not matching the claimed optimal decision in quadrants I and III and switching to the claimed optimal decision yields a lower FDR and a higher objective function. 
The decisions in quadrants II and IV are together rank-optimal. To prove that, among and across quadrants II and IV, we show that not altering or respecting the base decision rules by ranking, i.e., to exclude first in quadrant II and to include first in quadrant IV, leads to a suboptimal procedure. In other words, switching from any final decision that does not match our proposed procedure to one that matches it will lead to a lower FDR and a higher objective function value.

We build the proof in steps. Define baseline decisions $\delta^b_i \equiv 1$ (inclusion) in groups I and IV, and $\delta^b_i \equiv 0$ (exclusion) in groups II and III. We apply the baseline decision for all tests $ i=1, \cdots, m$. Recall that we use the values and the weights to rank the hypotheses. The baseline decisions indicate whether their weights are negative. Now consider an update to the baseline decision rules $\pmb\delta^u(t)=\{\delta_i^u(t): i=1, \cdots, m\}$, where for a $t > 0$:

\[
\delta_i^u (t) =
  \begin{cases}
  1, & \text{   if   $value_i/weight_i \geq t$}, \\
  0, & \text{   if   $value_i/weight_i < t$}.
   \end{cases}
\]

Finally, define the final decisions as $\delta_i^* (t) := \delta_i^u (t)(1-\delta_i^b (t)) + (1-\delta_i^u (t))\delta_i^b (t)$. Note that for the hypotheses in quadrants I and III, the value-to-weight ratio is negative and, by definition, never updated. 

The final decisions will depend on a cutoff $t^*$, above which decisions are updated. To choose this cutoff $t^*$ along the ranking, denote the ranked hypotheses from quadrants II and IV as $H_{(r)}$, where $(r)$ indicates their rank. All hypotheses' corresponding costs and lfdr statistic will be denoted as $c_{(r)}$ and $lfdr_{(r)}$, respectively. The total available capacity is defined by $C_T := \sum_{i:lfdr_i < \alpha} (\alpha - lfdr_i) c_i$. Only groups I and IV contribute to this capacity. We define the cumulative capacities along the ranking for groups II and IV as $C(j):= \sum_{(r)=1}^j c_{(r)} |lfdr_{(r)} - \alpha|$. Along the ranking, choose $k := \max\{j: C(j) \leq C_T\}$. Here, $k$ is the rank of the hypotheses where the algorithm will stop updating the baseline decisions. These steps define the procedure.

In quadrants I and III, switching from the baseline decisions, i.e., choosing to exclude when it was included (group I) and to include when it was excluded (group III), leads to a lower capacity and value. Thus, regardless of the total available capacity, avoiding taking the baseline decisions in these groups is suboptimal.

For the remainder of the proof, we assume that the algorithm makes a random update decision (uses a randomization device) for the last hypothesis before stopping, such that it achieves the exact capacity overall in its final decisions. We emphasize that the randomization device is optional for the implementation of the procedure, but it assists in the proof.
We show that not following the ranking order for groups II and IV when making update decisions is suboptimal. For the hypotheses beyond the algorithm's stopping rank, their baseline decision is not updated in the final decision because the randomization achieves the exact capacity. 
For example, suppose a hypothesis in group II (initially excluded by the baseline decision) is not included in the final decision by rank. In that case, it requires further capacity, but the ranked hypotheses have already exhausted the entire capacity. Similarly, excluding the hypothesis already included by the baseline decision (group IV) beyond the stopping rank reduces the total capacity. 

 Another possible change in the knapsack is to switch between hypotheses above the stopping rank and one below the stopping rank. This is suboptimal because swapping a decision outside the chosen set with decisions (or a fraction thereof) inside the selected set provides a lower total value since the hypotheses have already been ranked by their value-to-weight ratio. \hfill\BlackBox


\subsection{Second-Order Mean Correction}
While we approximate $\ln \hat{p}_{1}/\hat{p}_{0}$ to a normal distribution as in \cite{katz1978obtaining}, we recommend doing a further second-order mean correction because our simulations show that a standard normal as derived by \citep{katz1978obtaining} can be made more accurate. 

Let $0 < p_1 < 1$, then define $f(p) = \ln p$, and by the second order approximation $f(p) \approx f(p_1) + (p-p_1)f'(p_1) + (p-p_1)^2f''(p_1)/2$. Plugging in $p = \hat{p}_1$, we get $\ln \hat{p}_1 \approx \ln p_1 + (\hat{p}_1 - p_1)f'(p_1) + (\hat{p}_1-p_1)^2f''(p_1)/2$. Where $f''(p_1) = -1/p_1^2$. Thus the second order term is $-(1/2) (\hat{p}_1/p_1-1)^2$ which in expectation equals $(-1/2) Var(\hat{p}_1/p_1)$. The result for $0 < p_0 < 1$ is similar, i.e., $E \ln (\hat{p}_0/p_0) \approx -(1/2) Var(\hat{p}_0/p_0)$ and not zero as is usually assumed. 

The correction procedure would add estimates of $(1/2) Var(\hat{p}_1/p_1)$ and $(1/2) Var(\hat{p}_0/p_0)$ when computing the lfdr.
Our simulations found that this correction approximates the logarithm function better.\footnote{A similar correction to the variance estimate can be applied. It will require further derivations, but will likely not impact precision much.}


\subsection{Properties of the Data-Driven Procedure}

In practice, the lfdr statistic has to be estimated using the data. There are a few approaches to estimating the lfdr statistic:~by \cite{efron2015package} or \cite{qvalue}, \cite{sun2007oracle} via \cite{jin2007estimating}, and Gaussian mixture models (see \cite{fraley2002model} for a general exposition). Because the lfdr is estimated, we prove that the data-driven approach asymptotically maintains the level of control of the FDR as the oracle approach. Later, we use numerical simulation to demonstrate the procedure's finite sample performance. 

\vspace{2mm}
\noindent
{\bf Proposition 3 (Asymptotic Validity.)} {\it The data-driven procedure asymptotically controls the FDR at the desired level.} 


\vspace{2mm}
\noindent
{\bf Proof}. Define the estimated weight of hypotheses $i$ as $\widehat{W}_i := (\widehat{lfdr}_i - \alpha) \cdot c_i$ and the ratio of value to weight $\widehat{R}_i := \Pi_i \cdot \{ \widehat{lfdr}(h_i)/\widehat{lfdr}(h_i + \hat{\sigma}_i) - 1 \} /\widehat{W}_i$. To simplify the proof, we assume w.l.o.g.~that $c_i \equiv 1$ and $\Pi_i \equiv 1$. 

For every threshold $t \in [0, 1]$, we calculate the capacity of items included in the knapsack by the algorithm, averaged over all hypotheses. 
The knapsack includes all the hypotheses with positive weights and positive values higher than the threshold. For hypotheses with negative weights, the knapsack will consist of those with non-negative values or hypotheses with a value-to-weight ratio lower than the threshold. 
We define this capacity as $$\widehat{Q}(t) := \frac{1}{m} \sum_{i=1}^m \widehat{W}_i \{ 1_{\widehat{W}_i \geq 0} 1_{\widehat{R}_i \geq t} + (1-1_{\widehat{W}_i \geq 0})(1-1_{\widehat{R}_i \geq t}) \}.$$

By construction, the data-driven procedure ensures $\mathbb{E}_{\widehat{lfdr}}\{ \sum_{i \in \mathcal{R}_K} (1-\theta_i)/K\} \leq \alpha$ where
$\mathcal{R}_K$ is the set of rejected hypotheses and $K$ is the number of rejections. This guarantee only holds for the estimated lfdrs, and now we show that it will hold for $$\mathbb{E}_{\theta, data} \{ \sum_{i \in \mathcal{R}_K} (1-\theta_i)/K\} \leq \alpha.$$

First, we note that
\[
\sum_{i \in \mathcal{R}_K} (1-ldfr_i)/K = \sum_{i \in \mathcal{R}_K} (1-\widehat{ldfr}_i)/K + \sum_{i \in \mathcal{R}_K} (\widehat{ldfr}_i -lfdr_i)/K.
\]
To complete the proof we need to show that $E( \sum_{i \in \mathcal{R}_K} (\widehat{ldfr}_i - lfdr_i)/K) = o(1)$. The quantity within the expectation is bounded by 1. Thus we need to show that $ \sum_{i \in \mathcal{R}_K} (\widehat{ldfr}_i-lfdr_i)/K = o_P(1)$. Note 
\[
|m^{-1} \sum_{i \in \mathcal{R}_K} (\widehat{ldfr}_i - lfdr_i)|  \leq  m^{-1} \sum_{i} |\widehat{ldfr}_i-lfdr_i| = o_P(1),
\]
which holds because $|\widehat{ldfr}_i-lfdr_i| = o_P(1)$.\footnote{
This assumption is similar to those in the results of \cite{basu2018weighted}, Appendix B.2.} 

This expression will induce a multiplication by $m/K$, which we prove is $O_P(1)$.
Define the data-driven threshold where the capacity runs out as $t^* := \inf \{ t \in [0,1]: \widehat{Q}(t) \leq 0 \}$. We can rewrite $m/K$ as 

\[
\frac{m}{\sum_i \{1_{\widehat{W}_i \geq 0} 1_{\widehat{R}_i \geq t^*} + (1-1_{\widehat{W}_i \geq 0})(1-1_{\widehat{R}_i \geq t^*}) \}} \leq \frac{m}{\sum_i (1-1_{\widehat{W}_i \geq 0})(1-1_{\widehat{R}_i \geq 0})},
\]
which leads us to,
\[
\frac{m}{ \sum_i (1-1_{\widehat{W}_i \geq 0})(1-1_{\widehat{R}_i \geq 0})} = \frac{1}{E\{ lfdr(h+\widehat{\sigma}) < lfdr(h) < \alpha \} + o_P(1)} = O_P(1).
\]

In other words, we would like to guarantee that the probability of rejecting the hypotheses with negative weights and positive values has a positive mass, i.e., a lower-bounded constant. To make this claim, we bound the expression within the expectation by assuming that:
\[
P(lfdr(H+S) < lfdr(H) \leq \alpha) \geq \widetilde{p}_{\alpha},
\]
for some $\widetilde{p}_{\alpha} \in (0, 1]$ where $(H, S)$ is an i.i.d copy of $(H_i, S_i)$, the scaled $Z_i$ and the standard error. This condition must hold for any fixed (non-diminishing) nominal level $\alpha$ choice. This isn't a restrictive condition, as this occurs for large values of the scaled $Z_i$ in quadrant I, which happens with positive probability. \hfill\BlackBox


%

\section{Numerical Illustrations}

\subsection{Synthetic Data}
\label{sec:sims}
Using simulation, we compare our procedure to the one-sided no correction and two FDR control procedures:~adaptive BH \citep{benjamini2006adaptive}, and SC \citep{sun2007oracle}. Neither control procedures use the value of the lift in its objective. Because our method takes the lifts into account in the objective, we expect a significant increase in total lifts with the same level of the FDR when using our method compared to the other two. \cite{berman2022false} inspired part of our setting. However, we vary the baseline (aka null) conversion rates and fix the difference from the non-null conversion rates (the effect size). Hence, experiments with lower baseline conversion rates generate larger lifts. All simulation results we report are the average of 400 replications. We assume each experiment is an A/B test with one baseline (control) and one treatment variation. 

We report three different simulations. Each simulation has 2,000 experiments, and each arm in the A/B test has 5,000 visitors.
The baseline conversion rates follow a beta distribution:~uniform Beta(1, 1), mildly left-skewed Beta(1, 0.5), and highly left-skewed Beta (1, 0.25). The treatment effect, i.e., the difference between the non-null and baseline conversion rates, is zero with a probability of 0.8 and positive or negative 0.01 with a probability of 0.1 each.

The desirable level of FDR control $\alpha$ is 0.05. However, the insights do not change for different levels of $\alpha$, i.e., more conservative (0.01) or more liberal (0.1).

An essential part of our setup is to address experiments with heterogeneous baseline profits ($\Pi_i$). For every experiment, we draw $\Pi_i$ from gamma distributions with parameters (1, 1), (1/4, 1/4), and (1/9, 1/9). This results in a 3x3 design (3 baseline Beta distributions and 3 profit weight Gamma distributions).\footnote{We report only a selected share of these combinations as the insights do not change.}
These Gamma distributions have a longer right tail with their means fixed at one, but the scale varies from one to three. 

\begin{figure}
\begin{center}
\includegraphics[width=0.8\textwidth]{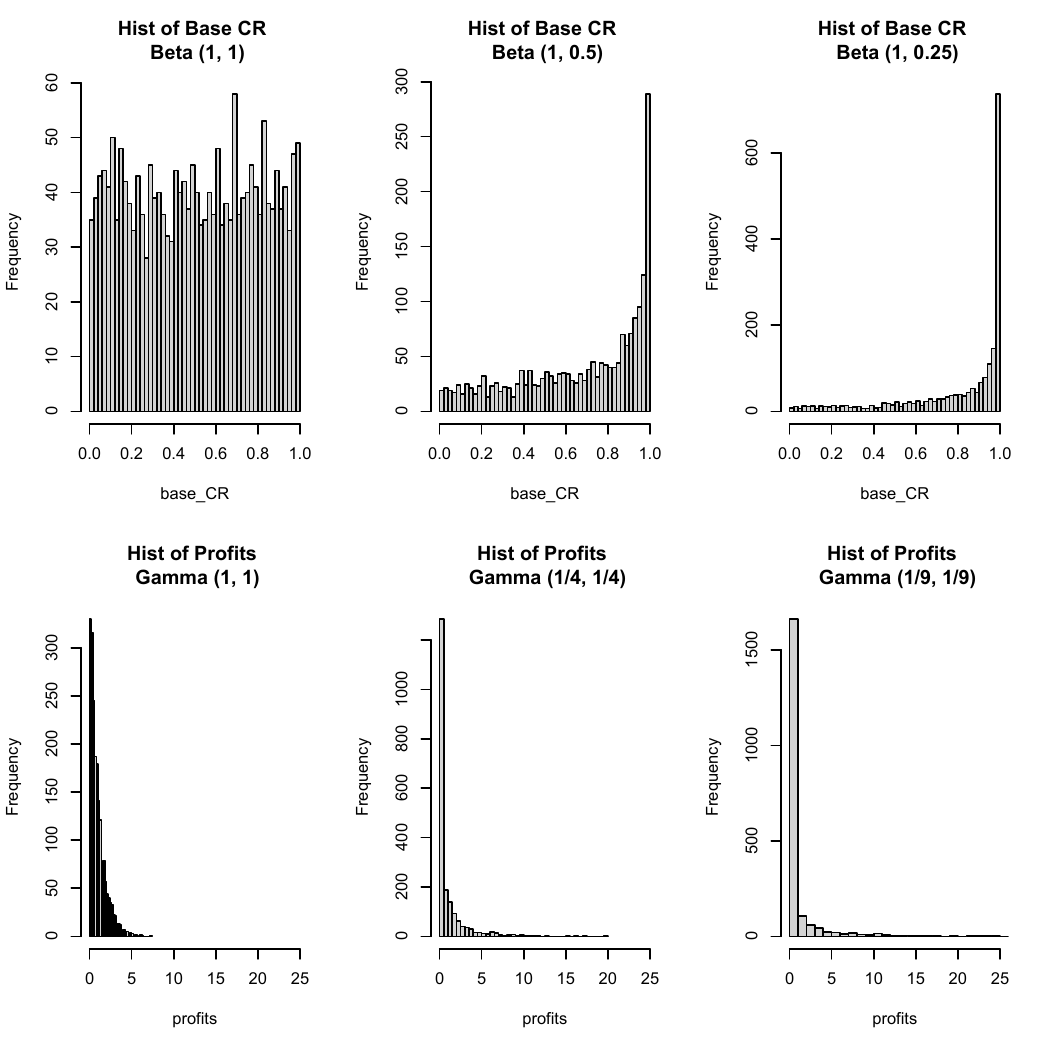}

\end{center}

\caption{Graphical illustration of the baseline conversion rates and profits.}
\label{fig:dist}
The figure displays the realized baseline conversion rates and profits of 2,000 experiments from the following Beta and Gamma distributions. [Upper panel, left to right] Baseline conversion rates are distributed Beta with parameters (1, 1), (1, 0.5), and (1, 0.25), respectively. Note that the top left plot corresponds to a realization from the uniform distribution on [0, 1]. Moving further to the right makes the distribution increasingly left-skewed. [Lower panel, left to right] Baseline profit distributions are Gamma with parameters (1, 1), (1/4, 1/4), and (1/9, 1/9), respectively. Note that the standard deviation increases from one to three while keeping the mean profits at one for all the distributions. 

\end{figure}

Figure \ref{fig:dist} provides a sample realization of the baseline conversion rates and profits. The setup with Gamma(1, 1) corresponds closely to the unweighted cases. We expect a higher objective function or total profit in the skewed cases from incorporating the weights. Because of the weighted nature of the heterogeneity, we further compare the performance of our procedure with the weighted FDR control procedure of \cite{basu2018weighted}, which we denote BCDS. We expect our procedure to perform better because the total lifts are taken into account directly when controlling for the FDR.

Table \ref{tb:fdr-comp} presents the simulation results that compare our procedure to the ones described. Understandably, the uncorrected procedure has a very high directional FDR (40\%-50\%) and is not recommended.

The adaptive BH method we used extends the original method by adjusting for an estimated null proportion $\pi_0$. The adaptive BH performs comparably to the SC method, which uses the lfdr as the main statistic and not p-values like BH. The BH, SC, BCDS, and the proposed procedures all control the FDR at the desired level of 0.05. The BH and the SC methods have comparable profits. Compared to them, BCDS and RBL have relatively higher profits in the range of 20-40\%. When the profit distribution's scale parameter decreases to 1, the difference between BH, SC, and BCDS vanishes. However, the RBL procedure continues to have a relative 5-10\% higher profit compared to its nearest competitor, BCDS, which still uses the weights but does not use the lifts in the objective function. 
In another simulation (not reported here), decreasing the effect size from 0.01 to 0.001 shows a relative profit gain of 20\% compared to BCDS. That is, if the hypothesis test problem is harder (smaller effects), the gain of the procedure is higher. 

The statistical power percentage refers to the proportion of correct directional hypotheses identified. We observe that the BH or SC methods have higher power at the cost of lower weighted discoveries or weighted lifts than BCDS or the proposed RBL method. 

To summarize, Table \ref{tb:fdr-comp} shows that:~(i) The uncorrected method is severely under-performing in terms of FDR; (ii) when the baseline profits are heterogeneous, weight-based FDR control methods have desirable FDR properties; and (iii) if one cares more about profits, RBL can be used to incorporate the lifts and achieve higher gains than BCDS.

\begin{boxedtable}{Comparison of different procedures}{tb:fdr-comp}
\begin{ctabular}{l l rrrrrrr} 
\midrule
& & Uncorrected & BH &  SC & BCDS & RBL \\  
\midrule

$\theta_{\gamma} = 1$ & FDR & 0.499 & 0.043 & 0.043 & 0.050 & 0.050 \\ 
$\beta = 1$& Power (\%) & 40.491 & 7.379 & 7.333 & 6.781 & 6.229 \\ 
& Profit (\%) & 72.003 & 44.139 & 44.035 & 44.977 & 46.988 \\ 
\midrule 
$\theta_{\gamma} = 2$ & FDR & 0.497 & 0.045 & 0.045 & 0.047 & 0.046 \\ 
$\beta = 1$& Power (\%) & 40.447 & 7.228 & 7.133 & 5.895 & 5.659 \\ 
& Profit (\%) & 66.802 & 34.933 & 34.757 & 38.459 & 40.796 \\  
\midrule 
 $\theta_{\gamma} = 3$& FDR & 0.500 & 0.041 & 0.043 & 0.050 & 0.047 \\ 
$\beta = 1$& Power (\%) & 40.360 & 7.304 & 7.292 & 5.686 & 5.557 \\  
& Profit (\%) & 65.259 & 33.575 & 33.501 & 38.361 & 41.279 \\ 
\midrule
$\theta_{\gamma} = 3$ & FDR & 0.446 & 0.041 & 0.042 & 0.047 & 0.050 \\ 
$\beta = 0.5$& Power (\%) & 50.310 & 19.250 & 19.279 & 15.510 & 14.693 \\ 
& Profit (\%) & 59.909 & 29.255 & 29.245 & 37.166 & 40.167 \\ 
\midrule 
$\theta_{\gamma} = 3$ & FDR & 0.406 & 0.041 & 0.042 & 0.043 & 0.048 \\ 
$\beta = 0.25$& Power (\%) & 58.452 & 30.356 & 30.379 & 25.008 & 23.674 \\  
& Profit (\%) & 57.394 & 29.718 & 30.318 & 39.453 & 42.406 \\ 
\midrule 
$\theta_{\gamma} = 2$ & FDR & 0.403 & 0.039 & 0.040 & 0.044 & 0.050 \\ 
$\beta = 0.25$& Power (\%) & 58.861 & 30.186 & 30.311 & 26.241 & 24.207 \\  
& Profit (\%) & 60.743 & 32.839 & 32.868 & 38.756 & 43.829 \\ 
\midrule 
$\theta_{\gamma} = 1$ & FDR & 0.403 & 0.038 & 0.039 & 0.042 & 0.051 \\ 
$\beta = 0.25$& Power (\%) & 58.873 & 30.374 & 30.507 & 28.780 & 25.604 \\  
& Profit (\%) & 64.967 & 39.255 & 39.524 & 41.694 & 46.129 \\ 
& \textbf{Net Profit} & \textbf{2.880} & \textbf{4.109} & \textbf{4.115} & \textbf{4.276} & \textbf{4.618} \\
& \textbf{Lifts/Rej.} & \textbf{0.035} & \textbf{0.094} & \textbf{0.093} & \textbf{0.101} & \textbf{0.115} \\ 
\midrule 
\end{ctabular}

\tablexplain{The table compares the performance of the RBL procedure relative to the Uncorrected, BH, SC, and BCDS. The scale parameter for the gamma distribution for the profits is varied in $\theta_{\gamma} \in \{1, 2, 3\}$. In contrast, the shape parameter for the beta distribution for generating the baseline conversion rates is varied in $\beta \in \{1, 0.5, 0.25\}$. Each simulation considers 2000 tests. We repeat the exercise for 400 simulations. We implement the FDR control methods at a nominal $\alpha = 0.05$. For net profit calculations, we choose the mean cost-to-profit ratio at 2\%.} 
\end{boxedtable}

A higher percentage of power or profit for uncorrected tests is expected. In cases when it is hard to distinguish false from true discoveries, higher false rejections often mean more true rejections, leading to higher power. The argument is similar for the profit, as true rejections will have a positive lift.
We also report the net profit (i.e., profit after deducting the weighted costs) for the last simulation instance to give a sense of the economic benefit of incorporating FDR-based metrics. For illustrative purposes, we chose a mean cost-to-profit ratio of 2\%. Although the method does not optimize the net profit directly, we see that RBL achieves a higher net profit than all other methods, improving the uncorrected method by more than 60\% and by about 8-12\% from the BH and BCDS methods.

In the empirical application, we cannot observe the true lifts and hence cannot compute the profit. However, we can compute the observed average lift per rejection.
Assuming the FDR methods we compare to control the FDR at the same level, a better algorithm will generate fewer rejections or reject hypotheses with higher lifts. Hence, although in the empirical application, we will use lift estimates, which are noisy measurements of the true lifts, a higher lift per rejection rate likely indicates that the algorithm will generate a higher profit. This pattern is evident in Table \ref{tb:fdr-comp}, where we observe that a higher net profit also implies a higher average lift per rejection.

\subsection{Application:~Optimizely Experiments}
\label{sec:app}






The data includes results from 2,766 experiments on the Optimizely platform starting in April 2014 \citep{berman2022false}. The experiments were all website experiments conducted by more than 1,000 experimenting companies. Every experiment had one control (baseline) condition and one or more variations of treatments. In total, there are results from 4,965 variations in our data. Because the lifts in experiments with many variations and one baseline are possibly correlated, we include non-baseline variations selected randomly with a probability proportional to the inverse of the number of variations in each experiment. 
This random selection may contain a mild dependency and gives us about 2,700 experiments to analyze. For each experiment, we have data about the total number of unique online visitors exposed to a specific variation of a webpage (corresponding to  $n_{i0}$ and $n_{i1}$) and the total number of clicks on that page, which is the dependent variable in the experiment. These values allow us to compute the conversion rate (in terms of clicks) of the baseline control and the non-baseline variation. These conversion rates are used to compute lifts, z-scores, and p-values for the lifts. 

Similar to our synthetic data analysis, we compare the RBL method with the Uncorrected, BH, SC, and BCDS methods. In this case, the ground truth (true population level conversion rates) is unknown, so we cannot report the FDR, power, or profit. We report the number of rejections for each method at different nominal level choices of 0.01, 0.05, and 0.10. In this dataset, there is no heterogeneity in baseline profit per conversion. Hence, we initially do analyses without heterogeneity in baseline profits, but then extend them by generating heterogeneous profits from the gamma distribution as before. 

We cannot compute the net profit in this data-driven setup since the true number of false discoveries is unknown. However, we can impute the total weighted or unweighted (by baseline profits) observed lift, which, while not accurate, provides an estimate of the true lift. 
To evaluate the performance of our method in comparison to the others, we use the observed average lift per rejection. 

\begin{boxedtable}{Performance in Optimizely experiments}{tb:fdr-comp-app}
\begin{ctabular}{l l rrrrrrr} 
\midrule
& & Uncorrected & BH &  SC & BCDS & RBL \\  
\midrule

$\theta_{\gamma} = \text{NA}$ & \# Rejections & 179 & 111 & 111 & 112 & 105 \\ 
$\alpha = 0.01$& Est.~Lifts & 97.355 & 92.110 & 92.110 & 92.156 & 92.643 \\ 
& Lifts/Rej. & 0.544 & 0.830 & 0.830 & 0.823 & 0.882 \\ 
\midrule 
$\theta_{\gamma} = \text{NA}$ & \# Rejections & 324 & 142 & 145 & 145 & 133 \\ 
$\alpha = 0.05$& Est.~Lifts & 108.692 & 94.365 & 94.552 & 94.552 & 96.803 \\ 
& Lifts/Rej. & 0.335 & 0.665 & 0.652 & 0.652 & 0.728 \\  
\midrule 
 $\theta_{\gamma} = \text{NA}$& \# Rejections & 475 & 169 & 171 & 172 & 153 \\ 
$\alpha = 0.10$& Est.~Lifts & 117.133 & 96.971 & 97.083 & 97.115 & 100.402 \\  
& Lifts/Rej. & 0.247 & 0.574 & 0.568 & 0.565 & 0.656 \\ 
\midrule
$\theta_{\gamma} = 1$ & \# Rejections & 475 & 169 & 171 & 161 & 148 \\ 
$\alpha = 0.10$& Est.~Lifts & 116.707 & 96.194 & 96.367 & 99.182 & 104.662 \\ 
& Lifts/Rej. & 0.246 & 0.569 & 0.564 & 0.616 & 0.707 \\ 
\midrule 
$\theta_{\gamma} = 1$ & \# Rejections & 324 & 142 & 145 & 134 & 125 \\ 
$\alpha = 0.05$& Est.~Lifts & 108.009 & 94.214 & 94.325 & 96.155 & 99.520 \\  
& Lifts/Rej. & 0.333 & 0.663 & 0.651 & 0.718 & 0.796 \\ 
\midrule 
$\theta_{\gamma} = 2$ & \# Rejections & 324 & 142 & 145 & 127 & 123 \\ 
$\alpha = 0.05$& Est.~Lifts & 119.817 & 99.861 & 99.995 & 104.286 & 112.578 \\  
& Lifts/Rej. & 0.370 & 0.703 & 0.690 & 0.821 & 0.915 \\ 
\midrule 
\end{ctabular}

\tablexplain{The table compares the performance of the RBL procedure to the Uncorrected, BH, SC, and BCDS methods. The scale parameter for the gamma distribution for the profits is varied in $\theta_{\gamma} \in \{\text{NA}, 1, 2\}$, where `NA' indicates the homogeneous baseline profit scenario. We implement the FDR control methods at a nominal $\alpha = \{0.01, 0.05, 0.10\}$. We set the random number generator seed for the selection of experiments and the profit weight generation to 2766.}
\end{boxedtable}

In Table \ref{tb:fdr-comp-app}, we observe that the Uncorrected method rejects many hypotheses, with an average observed lift of approximately 33\%. From our simulation study, we expect that 40\%-50\% of these rejections may be false positives. Practitioners often use an FDR control procedure such as the BH, SC, or BCDS methods to counter that. In both the simulation and application, we expect the performance of the BH and the SC methods to perform similarly, as they both guarantee FDR control at the chosen nominal level. Indeed, we observe this behavior in our application. The BCDS method is nearly identical in the homogeneous baseline profit scenario. In the heterogeneous (weighted) scenario, the BCDS method has higher weighted estimated lifts and a higher average lift per rejection. Our proposed RBL method achieves the highest estimated lifts and average lift per rejection in all scenarios. 

Note that the RBL method is designed to maximize the expected true lifts subject to FDR control, and while our data-driven results show the performance in outcomes for estimated lifts, they potentially have the limitation that this observation might not apply exactly the same to the true lifts in the specific application. The simulation results and the supporting theory complement our understanding of the overall performance of the RBL procedure.

\section{Additional Related Work}
\label{sec:related}
This section positions our work within the extant literature on A/B testing and FDR control. \cite{azevedo2020b} considers a framework to screen innovations when the distribution of the underlying effects can be fat-tailed. They do not consider false discoveries, the costs of implementation, or the heterogeneity in baselines that our method addresses.  
\cite{azevedo2019empirical} provides an overview of classical and recent methods available in this context. Their summary primarily discusses estimation methods leading up to and including nonparametric empirical Bayes estimation strategies. Finally, \cite{gu2023invidious} discusses the ranking and selection of a pre-determined number of top interventions in a decision-theoretic framework, while our work considers the implementation of all `profitable' interventions, taking their costs into account.

Another consideration in the practical implementation of A/B is their online or sequential characteristics. \cite{johari2022always} implement and analyze elements of always valid p-values and confidence intervals under sequential testing. \cite{zhou2023building} discusses optimal boundaries for stopping rules. \cite{chiong2023minimax} minimize the maximum regret specifically for a multi-arm online experiment. \cite{feit2019test} analyzes the choice of sample size that maximizes cumulative profits. \cite{bartovs2022power} suggest decreasing the nominal level to reduce FDR, while \cite{schultzberg2024risk} discuss several business metrics implemented at Spotify. \cite{deng2016continuous} advocates Bayesian testing to justify continuous monitoring. \cite{gronau2019informed} also discusses Bayesian approaches to testing to incorporate prior expert beliefs on the competing hypotheses. These works focus on single test scenarios, not multiple or large-scale experimentation scenarios. We refer readers to \cite{larsen2024statistical} for an overview of statistical challenges in A/B testing.

\section{Discussion}
\label{sec:disc}
We provide an empirical Bayes approach (RBL) to ranking and selecting A/B tests by their expected lifts while satisfying a required level of FDR control. In contrast to other popular methods, the RBL method uses lifts as the primary metric and allows for heterogeneity in baseline conversion rates. This makes it suitable for scenarios with prior information about the baseline conversion rates (e.g., champion-challenger). Using this information about baseline conversion rates, the method achieves higher profit using the same FDR control level.
Our method also incorporates heterogeneous costs to naturally switch from the baseline to the variation. 

Our method faces the standard limitations of FDR control procedures. These procedures are often used as guidance to determine which hypotheses should be further researched using replication experiments. They do not guarantee that a hypothesis can be rejected with no error. 
Hence, after filtering out likely false discoveries, we recommend that experimenters replicate candidate hypotheses that pass the FDR control procedure before implementing them in production.

Future work can look into optimizing the experimental design while considering that an FDR procedure will be used during the analysis. For example, the number of variations and the sample size affect the cost of the experiment and the efficacy of the FDR control method.
How to best incorporate these considerations in our statistical decision-making framework has promising potential. 

{\cblue As mentioned in the introduction, our work focuses on experimentation programs \citep{sudijono2024optimizing}; however,} another potential avenue of inquiry is to analyze issues arising in online A/B testing due to optional stopping, peeking, or other post-hoc adjustments. Finally, because it is recommended to perform replication following an FDR control procedure (see, for example, \cite{berman2022false}), a future avenue for research is to understand how FDR control procedures and replications should be designed in conjunction to avoid the wastage of resources.





\acks{Pallavi Basu's SERB MATRICS award MTR/2022/000073 partially supports this research. Ron Berman acknowledges support from the ISB-Wharton Joint Research Initiative, Wharton's Dean Research Fund, The Mack Institute for Innovation Management, and the Wharton Analytics and AI Initiative.}

{\small
\singlespace
\bibliography{refs}
}


\newpage

{\cblue 

\appendix
\section{Analysis of Alternate Objective Formulation}
\label{app:alternate}
Recall that we aimed to solve:
$$
\max_{\delta_i \in \{0,1\}} \mathbb{E} \sum_{i} \Pi_i \ell_i \delta_i \text{~~subject to~~} \mathbb{E} \{ \sum (1-\theta_i) \delta_i c_i / \sum \delta_i c_i \} \leq \alpha.$$
A perhaps more straightforward formulation is to directly maximize the expected net profit:
$$
\max_{\delta_i \in \{0,1\}} \mathbb{E} \sum_{i} \{ \Pi_i \ell_i - c_i\} \delta_i .$$
This section discusses the statistical challenges of power and the validity of doing so. 

Even though this formulation is a compound decision problem, the oracle decisions are individual in nature owing to the separability of the objective function and independence of tests. The Bayes optimal solution in this setting is $\delta_i = 1$ if and only if $c_i/\Pi_i < E(\ell_i|\text{data})$. We discuss the merits of our formulation over this decision rule through a stylized simulation, with settings that closely match experimentation programs  \citep{sudijono2024optimizing} and our Optimizely data. 

Figure \ref{fig:plot_net_profit} illustrates the results. For a set seed (1), we generate 2000 samples of the log risk-ratio with 80\% coming from the null of $N(0, \cdot)$ and 20\% from a non-null of $N(0.1, \cdot)$. The variance is chosen to match the log risk-ratio asymptotic, assuming 5000 visitors and a conversion rate of 10\% for experiment versions A and B each. The cost-to-profit ratio takes the values of 2\% or 10\%, with the profit being fixed at five. Two oracle methods are compared, with our method being invariant to the scale of costs, and the Bayes optimal method for both values of the proportional cost are implemented. (Note that in the main text, all results in Section 5 are for the data-driven methodologies. Only this section compares the oracle methods and uses the same lfdr test statistics for both procedures for conceptual exploration.) 

We chose a nominal FDR level for our method of 0.1. Figure  \ref{fig:plot_net_profit} shows that the false discovery proportion is roughly maintained, and our method makes roughly 80 rejections, and about 70 of them are correct decisions. In contrast, we see that, depending on the cost-to-profit ratio, the Bayes optimal method makes either too many rejections, resulting in a large proportion (more than half) of them being false, or it rejects too few hypotheses, thereby also making the effort of running so many tests quite wasteful.\footnote{We report the results for one seed value:~they do not vary much for arbitrary choices of seeds.} Overall, if experimenters are interested in long-term learning from experimentation, simply optimizing the short-term net profit is not the best philosophy to rely on, as it can be very wasteful or too erroneous. In contrast, if the implementation of the challenger intervention is viewed as an investment, then controlling for the `wasted' proportion of total investment could be a reasonable constraint while maximizing the profit. 

Our empirical observation is similar to that in Figure 7a of \citep{sudijono2024optimizing} where the authors identify a range of p-value thresholds from liberal to conservative that are equivalent to the Bayes optimal decision rule, when compared to the nominal level, as the cost of implementation increases. We further emphasize that while calculating the cost of implementation is relatively straightforward, costs associated with human labor, as noted by the authors, costs associated with loss of correct learning, or cumulative costs associated with wrong knowledge are all expensive and uncertain for the firm. These costs cannot be easily measured, thereby leading to the additional difficulty of including them in a simplified objective function as discussed in this section. The idea of controlling the weighted FDR thus becomes more relevant owing to the more subtle interpretation of costs and false discoveries. 

\begin{figure}[ht]
\begin{center}
\includegraphics[scale=0.8]{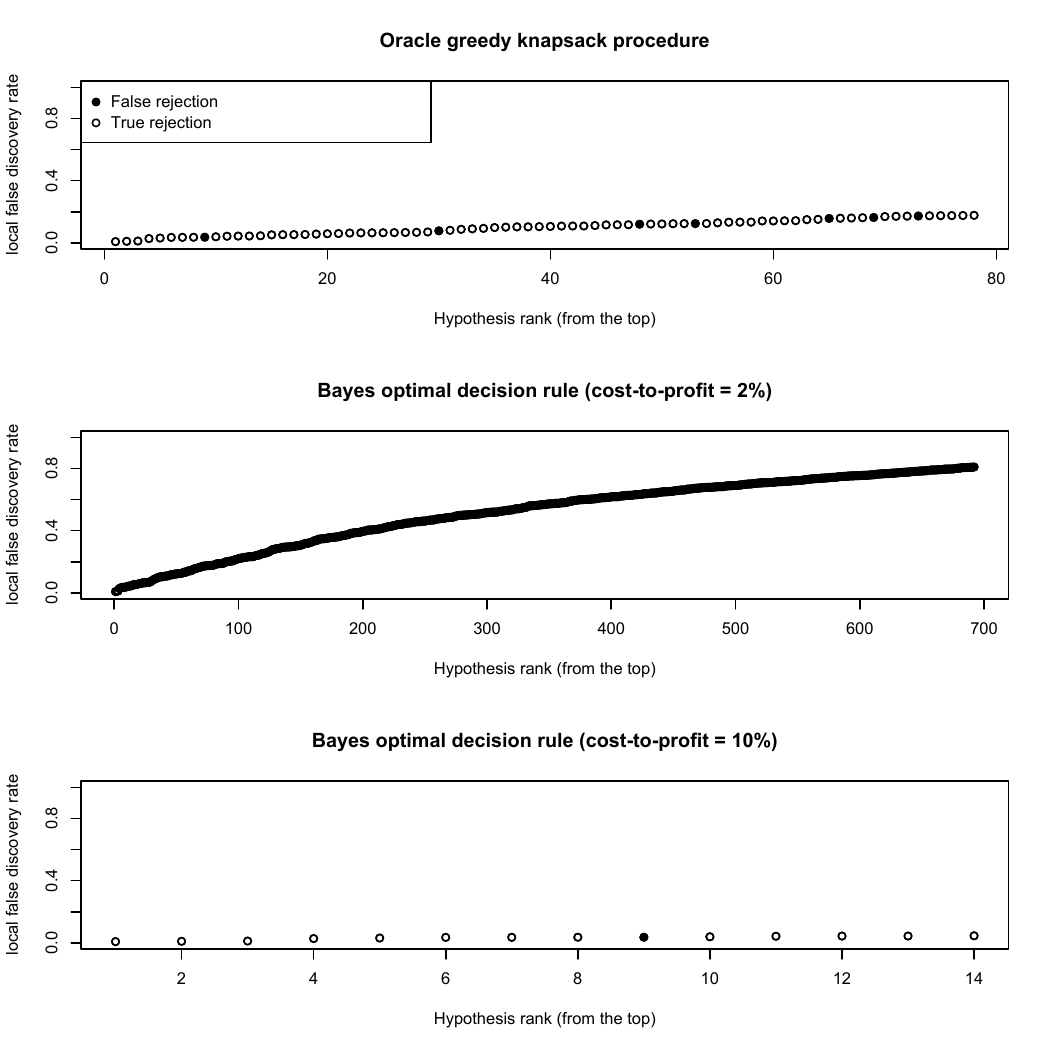}
\end{center}
\caption{Illustration of the rejection sets for our and alternate methods.}
\label{fig:plot_net_profit}
The figure displays the rejection set for our method when compared to the rejection sets of the Bayes optimal decision rule. The log risk-ratio process is assumed to be from a normal mixture with a 0.8 probability of being the null of $N(0, \cdot)$ and the non-null distribution is $N(0.1, \cdot)$. The variances are appropriately chosen. We choose $\alpha = 0.1$ for finding the rejection sets. We note that when the implementation cost-to-profit is a realistic 2\%, the Bayes optimal decision rule makes many more null rejections than is desirable. Conversely, when the implementation cost-to-profit ratio is higher, the Bayes optimal procedure makes hardly any rejections, making the A/B testing effort wasteful. 
\end{figure}

}

\end{document}